\documentclass[journal=jctcce,manuscript=article,layout=traditional]{achemso}

\usepackage{color}
\usepackage{cleveref}

\newcommand{\ie}{\emph{i.\,e.}~} 
\newcommand{\eg}{\emph{e.\,g.}~} 

\newcommand{\shole}{$\sigma$-hole}
\newcommand{\sholes}{$\sigma$-holes}
\newcommand{\relpf}{$rv_\mathrm{pf}$}
\newcommand{\Vmax}{$V_\mathrm{max}$}
\newcommand{\etal}{\emph{et al.}}
\newcommand{\brmet}{F$_3$CBr}
\newcommand{\imet}{F$_3$CI}

\title{Assessment of Scalar Relativistic Effects on Halogen Bonding and $\sigma$-Hole Properties}

\author{Michal H. Kol\'a\v r}
\email{michal@mhko.science}
\affiliation{Department of Physical Chemistry, University of Chemistry and Technology, Technick\'{a} 5, 16628 Prague, Czech Republic}

\author{Denisa Such\'a}
\affiliation{Department of Physical and Theoretical Chemistry, Faculty of Natural Sciences, Comenius University, Mlynsk\'a Dolina, 842 15 Bratislava, Slovakia}

\author{Michal Pito\v n\'ak}
\email{michal.pitonak@uniba.sk}
\affiliation{Department of Physical and Theoretical Chemistry, Faculty of Natural Sciences, Comenius University, Mlynsk\'a Dolina, 842 15 Bratislava, Slovakia}
\alsoaffiliation{Computing Center of the Slovak Academy of Sciences,
D\'ubravsk\'a cesta \v c. 9, 845 35 Bratislava, Slovakia}

\begin{document}

\singlespacing

\abstract

Halogen bond (X-bond) is a noncovalent interaction between a halogen atom and an electron donor. It is often rationalized by a region of the positive electrostatic potential on the halogen atom, so-called \shole. The X-bond strength increases with the atomic number of the halogen involved, thus for heavier halogens, relativistic effects become of concern. This poses a challenge for the quantum chemical description of X-bonded complexes. To quantify scalar relativistic effects (SREs) on the interaction energies and \shole ~properties, we have performed highly accurate coupled-cluster calculations at the complete basis set limit of several X-bonded complexes and their halogenated monomers. The SREs turned to be comparable in magnitude to the effect of basis set. The nonrelativistic calculations typically underestimate the attraction by up to 5\% or 23\% for brominated and iodinated complexes, respectively. Counter-intuitively, the electron densities at the bond critical points are larger for SRE-free calculations than for the relativistic ones. SREs yield smaller, flatter, and more positive \shole s. Finally, we highlight the importance of diffuse functions in the basis sets and provide quantitative arguments for using basis sets with pseudopotentials as an affordable alternative to a more rigorous Douglas-Kroll-Hess relativistic theory.

\maketitle

\section{Introduction}

Halogen bonds (X-bonds) are attractive interactions involving a halogen and a nucleophile. For the last two decades, they have grown to a notable noncovalent interaction with great potential in many areas of chemistry \cite{Cavallo16}. They have been utilized in material design to organize molecules to the desired crystalline forms \cite{Metrangolo08}. The supramolecular chemists benefit from the powerful ability of X-bonds to direct self-assembly in solution \cite{Gilday15, Brown16}. Similarly, the specificity of the interaction helped design new catalysts \cite{Carreras18} or modulate the action of existing ones \cite{Bulfield16, Carreras19}. In medicinal chemistry, the X-bonds are responsible for specific binding of halogenated ligands to biomolecules \cite{Wilcken13, Mendez17} which is reflected by the growing number of X-bonded complexes recognized.

What makes the X-bond particularly appealing for modern chemistry is its high directionality \cite{Kolar14} and the possibility to `tune' its strength. This is achieved either by exchanging the halogen atom involved in the interaction or by modifications in its chemical environment as suggested by theoretical work \cite{Riley11,Riley13}, although the practical use of the tuning is still not fully exploited \cite{Fanfrlik13}.

The rigorous definition of X-bond is rather recent \cite{Desiraju13}, despite the evidence of the puzzling attraction between halogens and electron-rich species stretching back to the 19th century \cite{Guthrie63}. There have been presented various views on X-bonds and their nature \cite{Eskandari10, Palusiak10, Wang14}. Perhaps the most successful rationalization of X-bonding was proposed by Politzer and co-workers who identified a region of positive electrostatic potential (ESP) located in the elongation of the R--X covalent bond \cite{Brinck92}. Later on, the region was called \shole \cite{Clark07} and its concept was expanded to other groups of the periodic table \cite{Murray09, Murray10}.

Substantial work has been done on the theory of X-bonding\cite{Kolar16}. Nevertheless, the accurate computational description of X-bonds is still challenging. The computations deal with all difficulties connected with an accurate description of noncovalent interactions in general \cite{Rezac16}. Let us name the basis set superposition error (BSSE), necessity to use diffuse atomic orbital (AO) basis functions, or the quality correlation effects treatment, should the dispersion energy play an important role. Naturally, the higher accuracy is required, the more computationally intensive the calculations are.

In addition to the aforementioned aspects, proper computational treatment of chemical systems containing heavy elements needs to consider relativistic effects. These are known to affect practically all atomic or molecular properties, and their importance tends to grow with the atomic mass\cite{Pyykko12}.

In the case of large (typically nonsymmetric) closed-shell systems, such as the ones studied in this work, relativity treatment can be reduced to so-called scalar (or spin-independent) relativistic effects (SREs) \cite{Almlof07,Ilias10}, if errors up to few (typically, tenths of) percents in interaction energies are tolerable \cite{Galland18}. This is a substantial simplification as the neglect of spin-orbit (SO) coupling makes calculations of extended molecular complexes computationally feasible. The use of a relativistic one-component Hamiltonian (instead of four- or two-component) approach comes with practically negligible overhead compared to nonrelativistic one.

In practice, two major routes to SREs treatment are pursued. Either it is the all-electron approach utilizing one-component relativistic Hamiltonian, such as Douglas-Kroll-Hess (DKH) formalism \cite{Najima12,Reiher12} (finite or infinite order), or the pseudopotential approach, in which the explicit core-electron interaction is replaced by a parametrized (often phenomenological) potential \cite{Schwerdtfeger11}. Despite its simplicity, this less rigorous approximation was shown to perform surprisingly well for a wide range of atomic and molecular properties and allows for even higher computational efficiency particularly at Hartree-Fock or Density Functional Theory level.

If we decompose a noncovalent interaction into electrostatic, induction, and dispersion forces, all of these contributions may be affected by SREs through dipole moments and dipole polarizabilities \cite{Ilias10}. SREs for X-bonds may be important, because common halogens, bromine and iodine, possess high polarizabilities \cite{Dejong98}, so in some instances, X-bonding may be dispersion driven \cite{Riley13b}. Holzer and Klopper studied \cite{Holzer17} dispersion bound rare-gas dimers and small molecular complexes and argued that SREs should be considered for atoms beyond Kr. 

Further in the context of noncovalent interactions, two studies reported contradictory results about relativistic effects in a gold-containing complex. While Anderson \etal~observed new topological features of the electron density when the relativistic effects were considered \cite{Anderson19}, Olejniczak \etal~argued that these are not significant and confirmed ``no relativity‐induced noncovalent interactions'' for the complex \cite{Olejniczak20}.

In the computational community, it was merely accepted that iodine SREs deserve special attention. Routinely, pseudopotentials are used in the chemical literature to treat SREs of iodine. Less frequently, higher-levels of relativistic theory (\eg SO coupling) are applied such as in the recent work \cite{Sarr20} focused on rare astatine X-bonded complexes \cite{Guo18}. The proper description of relativistic effects in astatine-containing molecules, as indicated also by other studies \cite{Hogan16, Hill13, Galland18}, is clearly more challenging compared to iodine and lighter halogens.

However, specific studies on the role of special relativity in X-bonding are rather sparse. Alkorta \etal ~studied the interactions of halogen diatomics with small Lewis bases \cite{Alkorta19} using relativistic pseudopotentials and a two-component ZORA Hamiltonian \cite{Lenthe93}. At the MP2 level with the triple-$\zeta$ basis set, they calculated various complex properties. While for the interaction energies the SREs were not assessed, a direct comparison of nonrelativistic and relativistic values was provided for NMR parameters. For instance, is was shown that the relativistic NMR chemical shifts are larger than their nonrelativistic counterparts \cite{Alkorta19}.

In this work, we perform quantum chemical calculations extrapolated to the complete basis set limit (CBS) using relativistic as well as nonrelativistic approaches. The use of a highly-accurate coupled cluster method allows us to dissect SREs from the effect of electron correlation treatment. Further the CBS extrapolation makes our conclusions independent of basis set size. Therefore we differentiate between the SREs and other effects in a comprehensive way.

We assess the impact of SREs on brominated and iodinated molecules and avoid any astatine, where non-scalar relativistic effects may play a role \cite{Galland18}, thus would go beyond our objectives. This is however only a minor limitation since the vast majority of studies focus on halogens up to iodine. Here, we calculate interaction energies of the X-bonded complexes, and on their electron density and ESP features. Understanding the role of SREs in the \shole ~properties should bring a transferable knowledge useful for applications regardless of the interacting partner of the halogen-containing molecules.

\section{Methods}

\subsection{Investigated Noncovalent Complexes and Basis Sets}

\begin{figure}[tb!]
    \centering
    \includegraphics{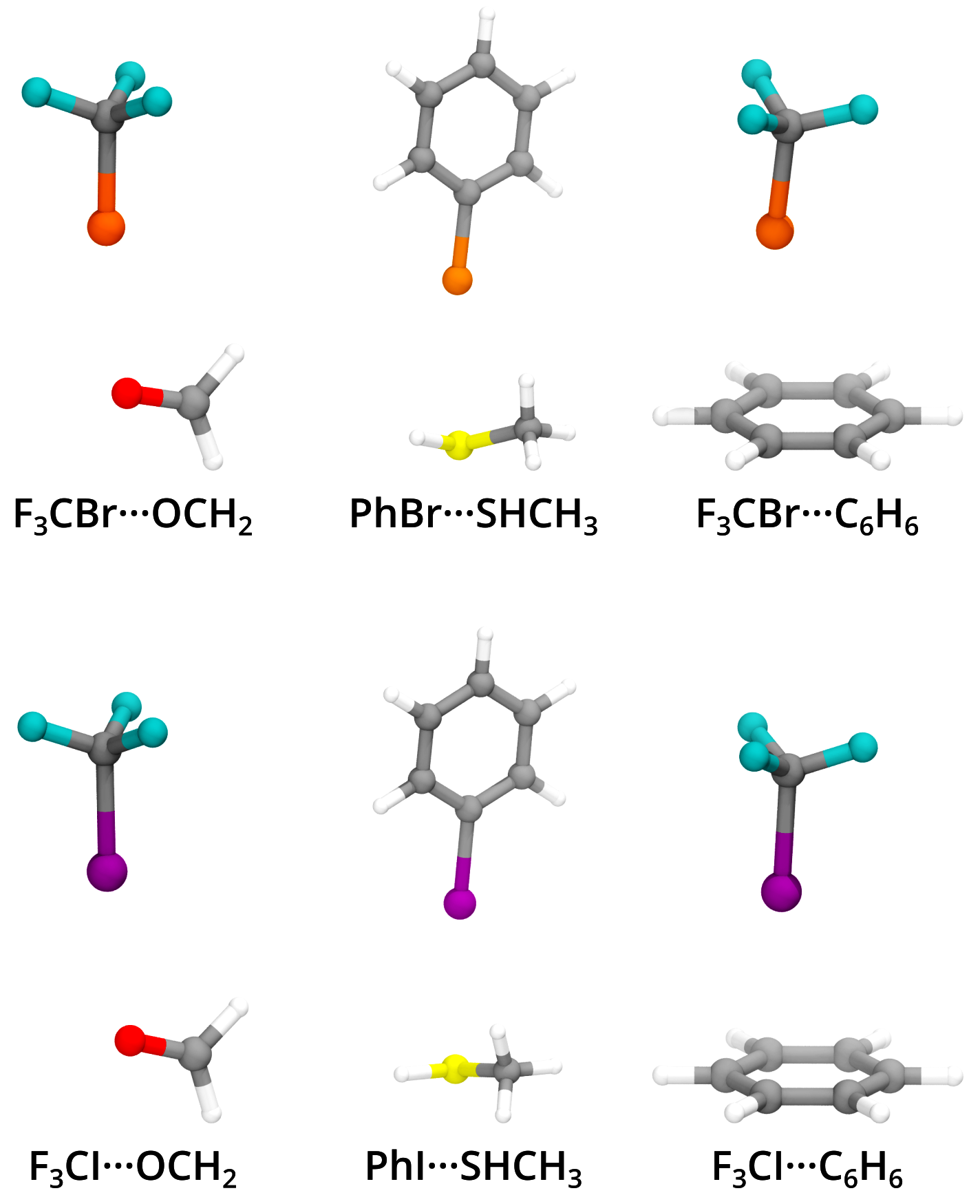}
    \caption{Geometries of the X-bonded complexes investigated. C gray, H white, F cyan, Br orange, I purple, S yellow.}
    \label{fig:geoms}
\end{figure}

The noncovalent complexes studied in this work were taken from the X40 data set of {\v{R}}ez{\'a}{\v{c}} \etal \cite{Rezac12}. The X40 data set contains benchmark interaction energies of noncovalent complexes, where halogen atoms participate in a variety of interaction motifs. In the original work, the X40 geometries of the complexes were optimized at the counterpoise-corrected\cite{Boys70} MP2/cc-pVTZ level, using pseudopotentials on halogen atoms. These geometries were used for our calculations without any further re-optimization. Here to probe diverse types of X-bonds, we selected a sample of six complexes from the X40, both aliphatic and aromatic (\Cref{fig:geoms}). Four of them form an X-bond with a lone electron pair:
\begin{itemize}
    \item tri\-fluoro\-bromo\-methane$\cdots$form\-aldehyde (F$_3$CBr$\cdots$OCH$_2$)
    \item tri\-fluoro\-iodo\-methane$\cdots$form\-aldehyde (F$_3$CI$\cdots$OCH$_2$)
    \item bromo\-benzene$\cdots$methane\-thiol (PheBr$\cdots$SHCH$_3$)
    \item iodo\-benzene$\cdots$methane\-thiol (PheI$\cdots$SHCH$_3$)
\end{itemize}
In the other two complexes, the X-bond directs towards a system of conjugated $\pi$-electrons:
\begin{itemize}
    \item tri\-fluorobromomethane$\cdots$benzene (F$_3$CBr$\cdots$C$_6$H$_6$)
    \item tri\-fluoroiodomethane$\cdots$benzene (F$_3$CI$\cdots$C$_6$H$_6$)
\end{itemize} 

The selection of basis sets\cite{Pritchard19,Schuchardt07a} suitable for this study was not straightforward. Ideally, we would have tested a series of nonrelativistic basis sets (with increasing cardinal number $\zeta$) and a complementary series, suitable for the side-by-side assessment of relativistic effects (using relativistic Hamiltonian and/or pseudopotential). The desire for a \emph{series} of basis sets is motivated by our intent to carry out basis set extrapolation of interaction energies towards the CBS limit, as they are known to converge rather slowly with the basis set size. Alternatively, we could choose explicitly correlated methods, but the availability of suitable (non)relativistic basis sets is even more limited.

The choice of basis sets for the bromine atom is somewhat wider than for iodine. For both halogen types, nonrelativistic XZP\cite{Camiletti08,Barros10,Canal05} (X=D, T, or Q, for double-$\zeta$, triple-$\zeta$, and quadruple-$\zeta$, respectively) and AXZP,\cite{Camiletti09a,Oliveira10a,Martins13a} as well as relativistic contractions for DKH Hamiltonian, XZP-DKH,\cite{Jorge09} are available. Nonrelativistic correlation-consistent basis sets (aug-)cc-pVXZ\cite{Wilson99} and DKH-optimized relativistic basis sets (aug-)cc-pVXZ-DK\cite{dkBasis} are available for bromine only, but relativistic pseudopotentials (aug-)cc-pVXZ-PP\cite{Petersen03} exist for both halogens. Another universal alternative for relativistic calculations is the ANO-RCC\cite{Roos04} basis set. Nonrelativistic ANO contractions were published as well, but not for heavier halogens than chlorine, so they were not used in this study.

\subsection{Interaction Energies}

Energies of complexes and their interacting subsystems were calculated using the coupled clusters method with iterative single- and double-excitations, corrected by perturbative inclusion of triple-excitations (CCSD(T)), wherever it was computationally feasible. The calculations in AQZP and aug-cc-pVQZ basis set of all complexes except for the smallest one -- F$_3$CBr$\cdots$OCH$_2$, could only be carried out at the second-order M{\o}ller-Plesset perturbation theory level (MP2), and the well-established focal-point scheme\cite{Jurecka02} was applied to obtain an estimate of the $E_Q$(CCSD(T)) energies.
\begin{equation}
\label{eq:focal}
    E_X(\mathrm{CCSD(T)}) =  E_X(\mathrm{MP2}) + \Delta E_{X-1}(\mathrm{CCSD(T)})
\end{equation}
where $\Delta E_{X-1}$(CCSD(T)) is the difference of CCSD(T) and MP2 energies in basis set with lower cardinality ``X -- 1''.

Energies at the CBS limit were estimated according to Halkier \etal \cite{HALKIER1998243} using the extrapolation formula
\begin{equation}
\Delta E_X = \Delta E_{CBS} + k.X^{-3}
\end{equation}
from triple- and quadruple-$\zeta$ basis sets, where X is 3 for triple- and 4 for quadruple-$\zeta$ basis, and k is a fitting constant. Interaction energies were corrected for BSSE using the counterpoise method \cite{Boys70}.

All calculations were carried out using NWChem 6.6 program package.\cite{Valiev10} The default threshold (of 10$^{-5}$ a.u.) for the elimination of linearly dependent AO basis functions had to be increased to achieve convergence of orbital optimization in some cases (Table S1). The third-order DKH approximation with cross-product integral terms\cite{Nakajima00} was used in all-electron relativistic calculations.

\subsection{Bond Critical Points}

For all complexes, the electron densities were exported from the NWChem as cube files and analyzed by the Topology Toolkit extension \cite{Tierny17} of the Paraview program package \cite{Ahrens05}. The electron densities were reported for the bond critical points (BCPs), \ie the (3, --1) critical points of the electron density \cite{Bader85}. The analysis was carried out for all basis sets and relativity treatments investigated with the additional aim to correlate the BCPs density $\rho_\mathrm{BCP}$ values with the interaction energies.

\subsection{Monomer Properties}

To characterize the halogen \shole s, electron densities and ESPs were analyzed at various {\it ab initio} levels. The supermolecular geometries of the monomers \brmet, \imet, PheBr, and PheI, as taken from the X40 data set, were used without further energy optimization. The analyses rested on the molecular surface defined as the isosurface of electron density of 0.001~e/bohr$^3$ \cite{Bader87}, which is a common threshold used in the context of X-bonding \cite{Kolar16}. 

The principal halogen covalently bound to a molecule is not a perfect sphere \cite{Sedlak15}, thus its size cannot be described by a single radius. Here, we use two radii to characterize the halogen size: i) the parallel radius $r_\parallel$ is a distance from the halogen to the electron density isosurface in the direction of C--X bond, ii) the perpendicular radius $r_\perp$ is equivalent distance, but in the direction perpendicular to the C--X bond. While the $r_\perp$ is uniquely defined for molecules with C$_\infty$ symmetry (\eg dihalogens), it must be explicitly specified in all other cases. Here for halogenbenzenes, we calculated $r_\perp$ as the arithmetic mean of two values: one $r_\perp$ lying in the aromatic plane, and the other perpendicular to it and to C--X bond at the same time. In the case of halomethanes, $r_\perp$ was calculated as the average of $r_\perp$ in the plane defined by X, carbon and one fluorine atom, and $r_\perp$ in the direction perpendicular to this plane.

The difference of the two radii represents the absolute value of polar flattening $v_{pf}$, as discussed in detail in ref. \citenum{Sedlak15}. Here, we focus on the ratio of $v_\mathrm{pf}$ and $r_\perp$ expressed in percents and denominated as the relative polar flattening $rv_\mathrm{pf} =100 \cdot (1 - \frac{r_\parallel}{ r_\perp})$. The \relpf ~describes how the halogen shape deviates from an ideal sphere.

The \shole ~magnitude \Vmax ~was calculated as the maximum of the ESP on the molecular surface in the extension of the C--X bond, as detailed in ref. \citenum{Kolar14}. All monomer calculations were based on the results of DPLOT and ESP modules of the NWChem package \cite{Valiev10}. Most of the analyses used NumPy \cite{VanDerWalt11} and Matplotlib \cite{Hunter07} Python libraries.

\section{Results and Discussion}
\label{sec:results}

\subsection{Characteristics of the Halogenated Monomers}

We describe how monomer characteristics vary with the SRE treatment, basis set type, and size. Because a wider range of basis sets is available for bromine than for iodine, we discuss the brominated monomers first, and later move to the iodinated ones. 

\begin{figure}[htb!]
\includegraphics[width=\textwidth]{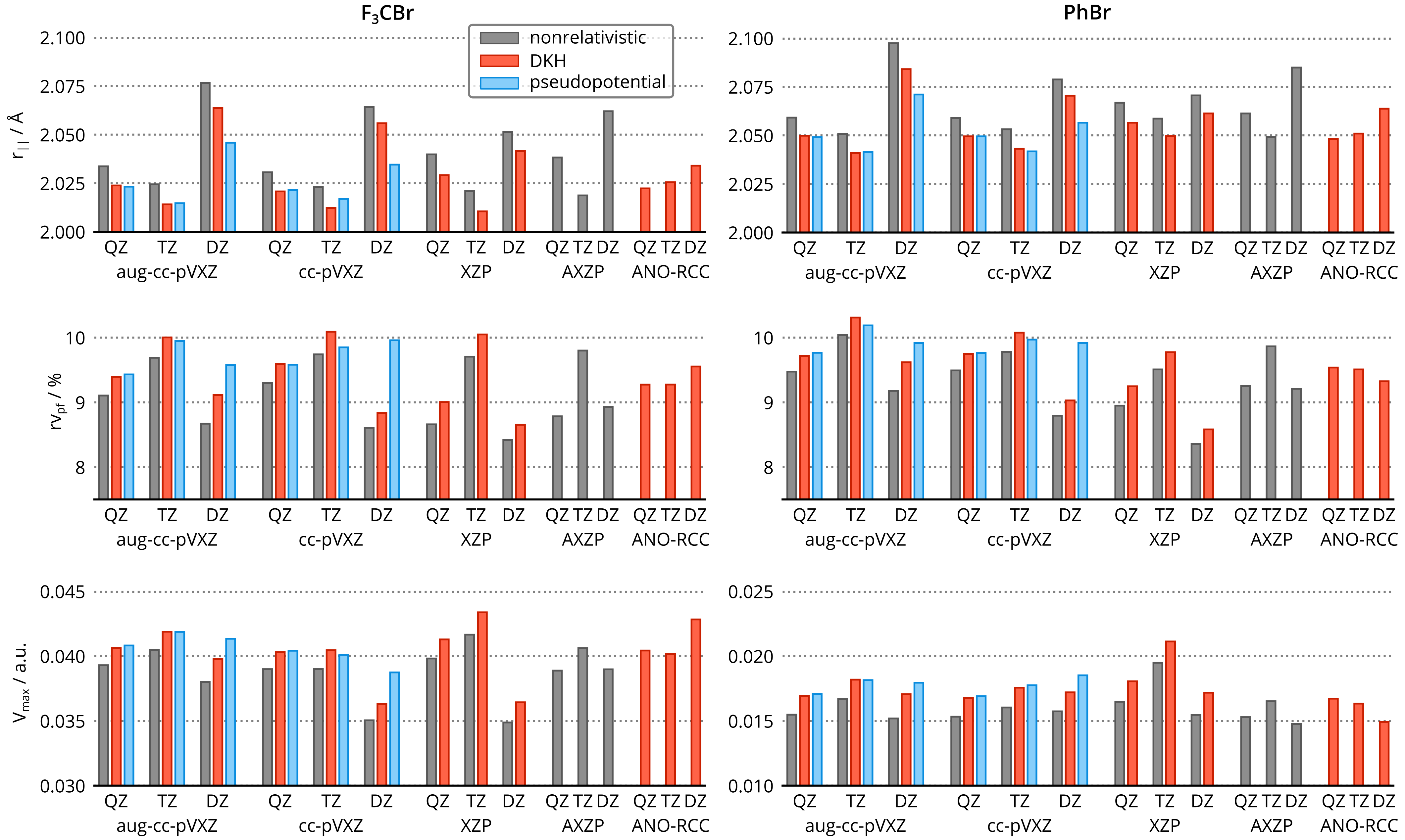}
\caption{The parallel halogen radius $r_\parallel$, relative polar flattening \relpf ~and \shole ~magnitude \Vmax ~of \brmet ~(left) and PhBr (right) calculated using various basis sets. Nonrelativistic (gray) and relativistic calculations using DKH (red) and pseudopotential (blue) approaches are compared, whenever possible.}
\label{fig:sigmaholesbr}
\end{figure}

\Cref{fig:sigmaholesbr} shows the $r_\parallel$, \relpf, and \Vmax  ~calculated for various basis sets and SRE treatment for the \brmet ~and PhBr. We assess the SREs on the monomer characteristics using values calculated in the largest available basis set, \ie the quadruple-$\zeta$ (QZ). The qualitative conclusions that bromine on the \brmet ~is smaller and more positive than the bromine on PhBr are true regardless of the inclusion or exclusion of SREs. Quantitatively on both monomers, the bromine appears smaller, flatter, and more positive when the SREs are taken into account (\Cref{fig:sigmaholesbr}). The differences between relativistic and nonrelativistic values are very small, however: about 0.01 \AA ~for $r_\parallel$, 0.2 per cent point for \relpf ~and 0.001 a.u. for \Vmax .

For the aug-cc-pVXZ and cc-pVXZ basis sets, a comparison of two relativistic approaches -- DKH and pseudopotentials -- is possible. In all monomer characteristics studied, the values are practically identical in the QZ basis set. Hence, it is beneficial to use the pseudopotential approach due to its lower computational demands compared to DKH, at least in one-electron approximations. A larger difference between DKH and pseudopotential treatment is observed for a double-$\zeta$ (DZ) basis set, especially for \relpf ~at the cc-pVDZ level. However, if we compare the DZ values with the reference QZ values with DKH treatment, the DZ with pseudopotential performs better than DZ-DKH, most likely due to a favorable compensation of errors.

The effect of basis-set size and basis-set type, which was reported previously for other basis sets and molecules \cite{Riley16, Kolar16}, is often larger than the SREs. Overall, it has a little effect on qualitative conclusions about bromine's $r_\parallel$, \relpf ~and \Vmax ~on the two monomers. Interestingly, low variations of $\sigma$-hole magnitude were obtained for the PhBr, which means that \Vmax ~of this system is less sensitive to basis set quality than \brmet.

\begin{figure}[htb!]
\includegraphics{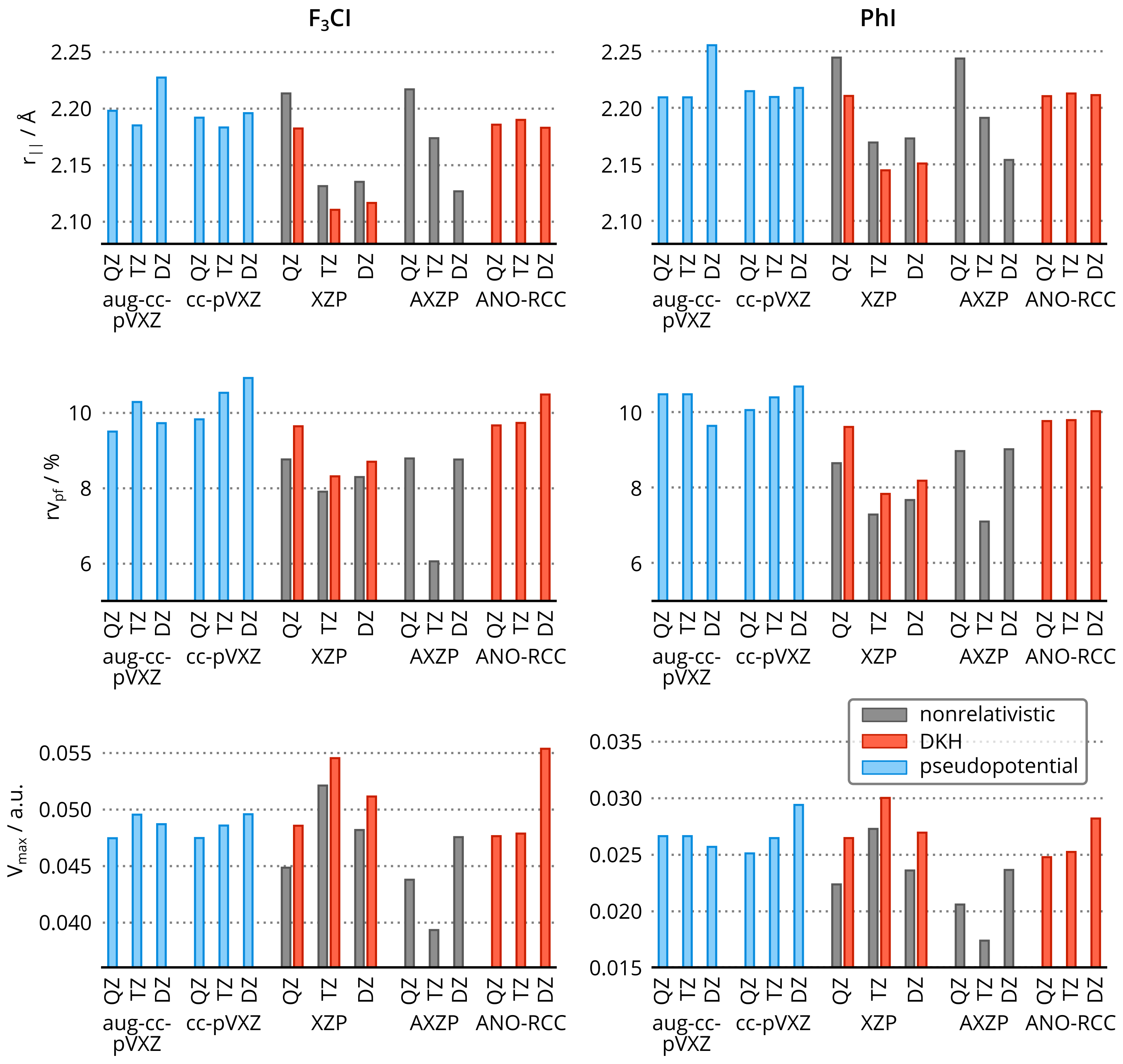}
\caption{The parallel radius $r_\parallel$, relative polar flattening \relpf ~and \shole ~magnitude \Vmax ~of \imet ~(left) and PhI (right) calculated using various basis sets. Nonrelativistic (gray) and relativistic calculations using DKH (red) and pseudopotential (blue) approaches are compared, whenever possible.}
\label{fig:sigmaholesi}
\end{figure}

For iodine, the nonrelativistic basis sets are less reasonable than for bromine due to iodine's higher mass. That is perhaps the reason why a narrower range of nonrelativistic basis sets is available. The $r_\parallel$, \relpf ~and \Vmax ~for the iodinated monomers \imet ~and PhI are in \Cref{fig:sigmaholesi}. A direct comparison between nonrelativistic and relativistic values is only possible for the XZP series. Alike bromine, taking SREs into account for iodine results in a smaller, flatter, and more positive atom compared to the nonrelativistic calculations. These qualitative conclusions are independent on basis set size. On the other hand, the values of the monomer characteristics often depend on the basis-set size more than on the relativity treatment. For instance, for \imet, the \Vmax ~difference between QZP and TZP is about twice as large as the difference between QZP and QZP-DKH (\Cref{fig:sigmaholesi}).

Following the chemical intuition, the extent of SREs is larger for iodine than for bromine for all three monomer characteristics. The differences between values with and without SREs are roughly 0.03 \AA ~of $r_\parallel$, about 1 per cent point of \relpf, and 0.004 a.u. of \Vmax ~(comparing QZP values).

\begin{figure}
    \centering
    \includegraphics{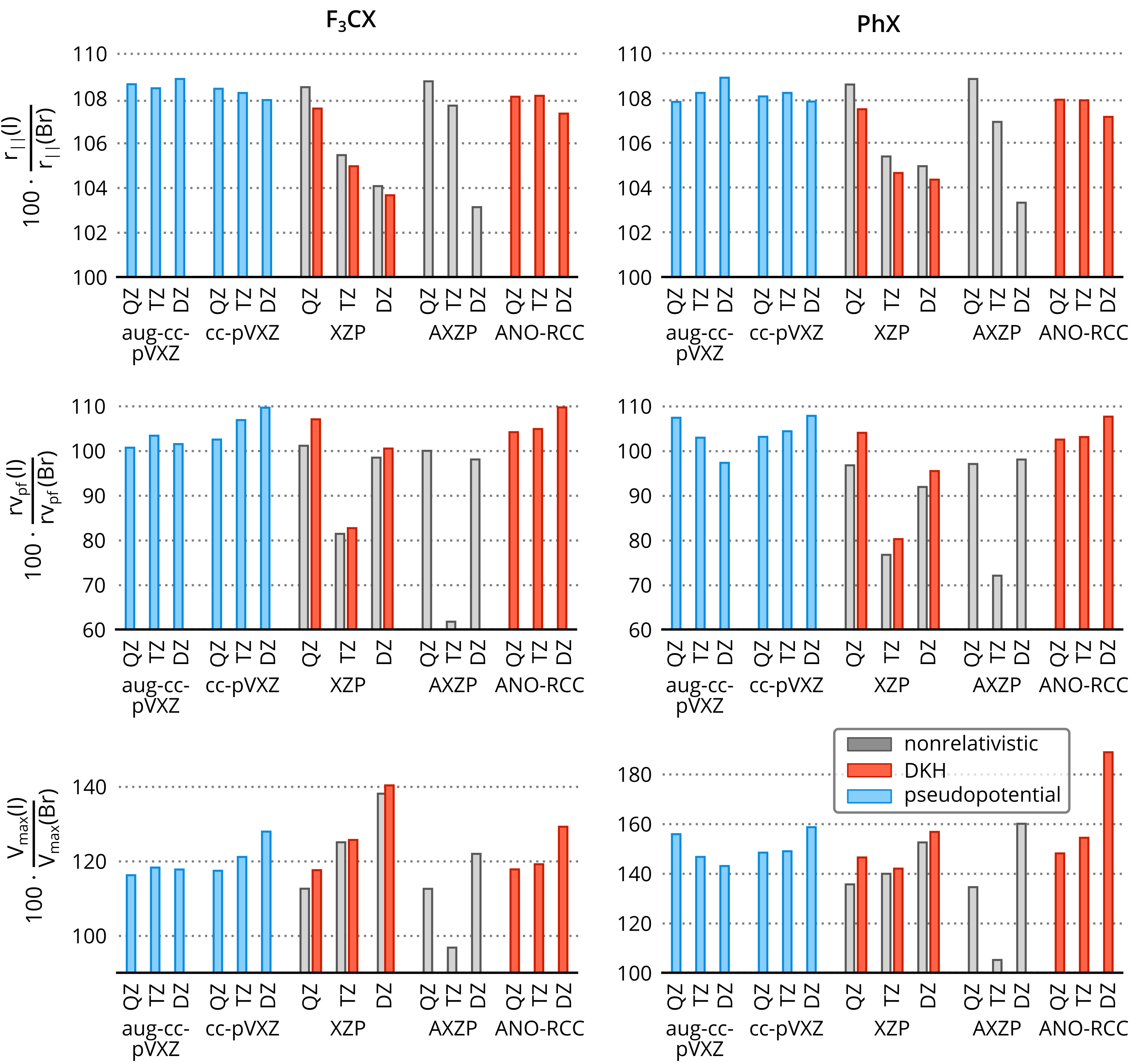}
    \caption{Ratios of the parallel radius $r_\parallel$, relative polar flattening \relpf ~and \shole ~magnitude \Vmax ~of iodinated and brominated monomers. Nonrelativistic (gray) and relativistic calculations using DKH (red) and pseudopotential (blue) approaches are compared, when possible.}
    \label{fig:i2br}
\end{figure}

Further, we investigate how much the iodine characteristics differ from the bromine characteristics and how these differences vary with the basis set and SREs. In other words, we want to know how SREs affect the relative comparison of the two halogens. \Cref{fig:i2br} shows the ratios $A$(I)/$A$(Br)$\cdot100$, where $A$ stands for $r_\parallel$, \relpf, or \Vmax. Note that we compare only those {\it ab initio} levels, which are available for iodine.

All calculations predict that iodine is larger than bromine (in a sense of $r_\parallel$) no matter what basis set or relativity treatment is used. The Dunning's (aug-)cc-pVXZ are remarkably consistent yielding iodine larger by 8\% than bromine. The same results are obtained for all ANO-RCC and QZ basis sets in general. There is a stronger dependence of $r_\parallel$ on the basis set size for XZP and AXZP series.

A notable qualitative effect of scalar relativity is observed for the iodine/bromine ratio of \relpf. While the relativistic calculations conclude that iodine is flatter than bromine on both trifluoromethyl as well as phenyl moieties ($rv_{pf}(\mathrm{I}) / rv_{pf}(\mathrm{Br}) > 1$), the nonrelativistic calculations predict the opposite. The effect of basis set is even larger: TZP and ATZP incorrectly predict iodine being less flattened than bromine by more than 25\%. More realistic results of DZP and ADZP are likely caused by the favorable compensation of errors.

The \shole ~magnitude is consistently predicted larger (more positive) for iodine compared to bromine, which goes in line with the notion in the X-bonding research community. The exception is the nonrelativistic ATZP, which puts the two halogens roughly on par and performs worse than the equivalent basis set of double-$\zeta$ size.

An important finding is that the pseudopotential calculations yield iodine/bromine ratios of monomer characteristics similar to the DKH values of both XZP and ANO-RCC series. This provides strong support for using pseudopotential calculations as a less demanding alternative to DKH treatment of SREs on halogenated molecules.

\subsection{Interaction energies}

Interaction energy should reflect the properties of interacting monomers, particularly if electrostatic energy is non-negligible, which is the case of the X-bond. Because the \Vmax ~values are larger when a relativistic approach is used, we would expect stronger intermolecular interaction for the relativistic calculations than for the nonrelativistic ones. On the other hand, X-bond is a superposition of many interaction energy components including \eg notable dispersion contribution and often it is not the exclusive interaction between the subsystems. The overall impact of SREs inclusion may thus be slightly smeared on the level of total interaction energies.

\begin{figure}[tb!]
\includegraphics{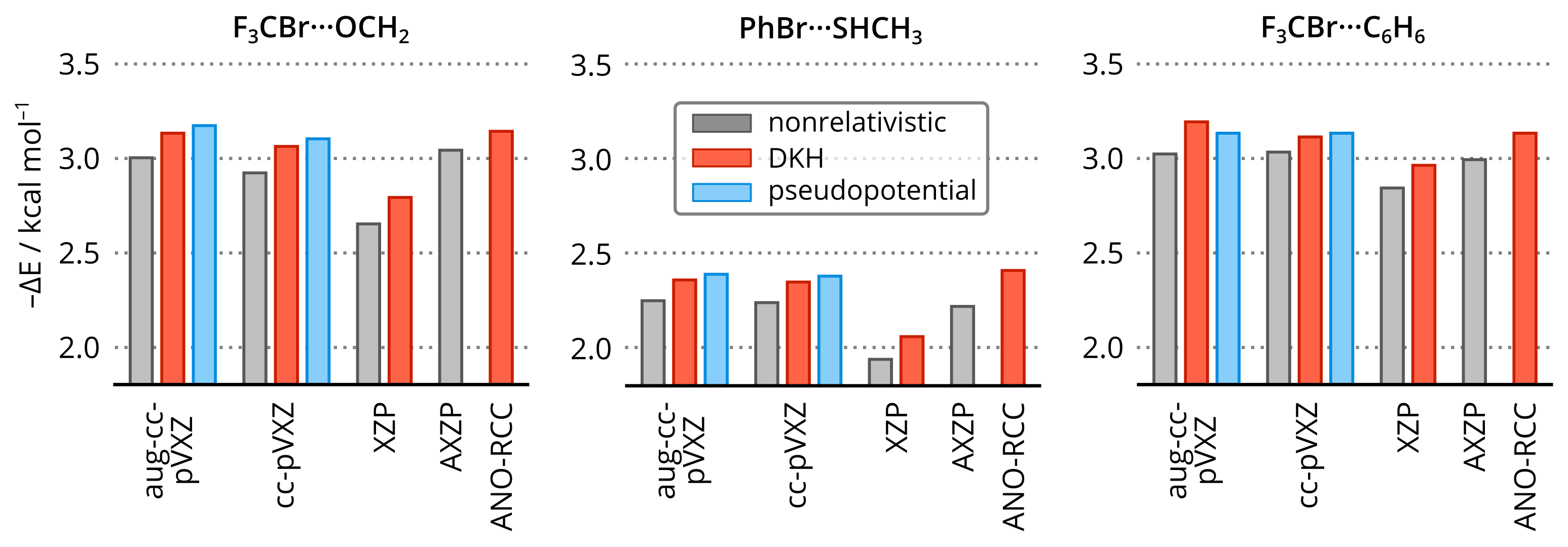}
\caption{The effect of basis-set type and size and of relativity treatment on interaction energies ($\Delta$E) of noncovalent complexes containing bromine. CP-corrected interaction energies extrapolated towards the CBS limit with nonrelativistic (gray), DKH-Hamiltonian optimized (red) and pseudopotential (blue) basis sets using frozen-core approximation are compared, whenever possible.}
\label{fig:inte_br}
\end{figure}

We inspect the brominated complexes, whose interaction energies obtained for several series of basis sets and SRE treatments are summarized in \Cref{fig:inte_br}. In general, the intermolecular interactions are slightly stronger when calculated at relativistic level, \ie the relativistic interaction energies are more negative than the nonrelativistic ones. The more positive relativistic \sholes ~agree with this finding. For brominated complexes, the SREs are in the order of about 5\% (or 0.1 kcal/mol). A comparison within (aug-)cc-pVXZ reveals that $\Delta E$ at pseudopotential level is a bit higher than at DKH level, but still, the (relativistic) pseudopotential approach is closer to the reference (relativistic) DKH value than the nonrelativistic calculation.

The basis set has a stronger impact on $\Delta E$ than the SREs. The XZP basis sets show the largest deviation from the rest of the results, regardless of the SREs inclusion, while AXZP and aug-cc-pVXZ basis sets deliver similar results. The XZP-DKH is by about 0.2--0.4 kcal/mol less negative than ANO-RCC (or aug-cc-pVXZ-DKH), which is about twice as much as the difference from nonrelativistic XZP values. Note that the basis-set size dependence of $\Delta E$ is damped by extrapolation towards CBS, so the variations are attributable to the basis set quality, \ie mostly sufficient presence of diffuse basis functions.

\begin{figure}[tb!]
\includegraphics{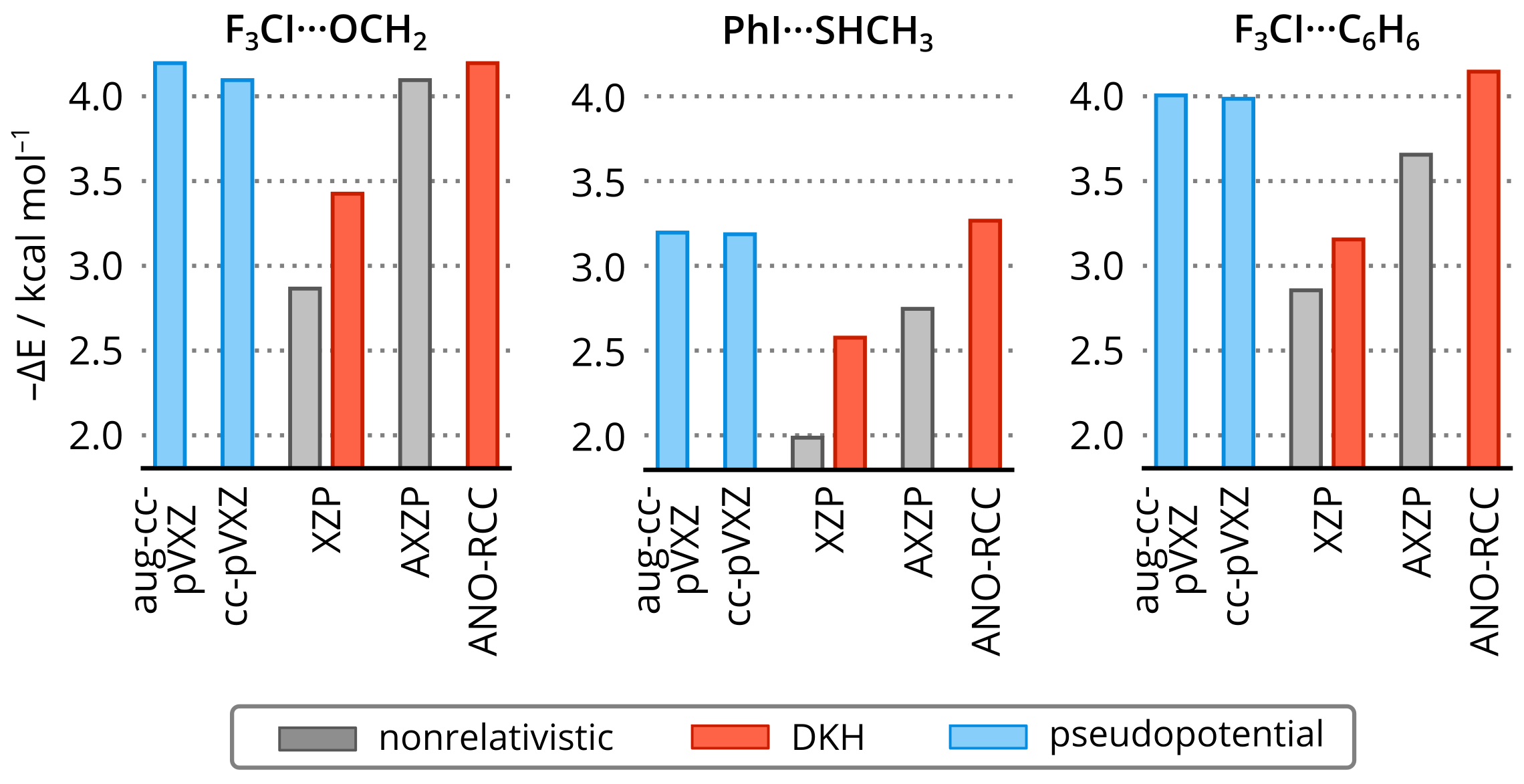}
\caption{The effect of basis set quality and relativity treatment on interaction energies ($\Delta$E) of noncovalent complexes containing iodine. CP-corrected interaction energies extrapolated towards the CBS limit with nonrelativistic (gray), DKH-Hamiltonian optimized (red) and pseudopotential (blue) basis sets using frozen-core approximation are compared, whenever possible.}
\label{fig:inte_i}
\end{figure}

For the iodinated complexes, the interaction energy plots are shown in \Cref{fig:inte_i}. Only the XZP basis set allows assessing the SREs directly. For all three complexes, the relativistic $\Delta E$ with XZP is always more negative than the nonrelativistic. Like in the bromine case, the interaction between the monomers is stronger when the SREs are taken into account. However, in complexes with iodine, the SREs are more pronounced and increase to almost 23\% (0.59 kcal/mol for PhI$\cdots$SHCH$_3$). 

Further, a more refined analysis is possible through $\Delta E$ values summarized in Tables S2 and S3. In the case of the smallest investigated system, F$_3$CBr$\cdots$OCH$_2$, we provide a full comparison of the extrapolated interaction energies obtained using the focal-point scheme (Eq. \ref{eq:focal}) and the `plain scheme', where CCSD(T) energies are calculated in quadruple-$\zeta$ basis sets. Deviations for the largest basis sets (AXZP and aug-cc-pVXZ(-DKH/-PP)), for which the focal-point scheme is used to obtain benchmark $\Delta$Es for larger complexes, are in the order of hundredths of kcal/mol. The deviations are more pronounces in smaller basis sets (up to about 0.1 kcal/mol or 4\%), which is understandable, if we take into account insufficient basis set saturation of the $\Delta E_{X-1}$(CCSD(T)) term.

The extrapolated $\Delta$Es for the bromine-containing species (Table S2) calculated in aug-cc-pVXZ-DKH basis set with SREs inclusion using DKH Hamiltonian are considered as the benchmark in this work. Should we accept the monotonous character of the $\Delta$E convergence with the basis set size (or cardinality), aug-cc-pVXZ-type basis sets exhibit slightly better saturation compared to ANO-RCC and AXZP basis sets. In the case of ANO-RCC basis sets this can be reasoned by a notably lower number of basis functions (by almost 40\%; 420, 726 and 724 for ANO-RCC quadruple-$\zeta$ contraction {\it vs.} 665, 1151 and 1169 for aug-cc-pVQZ for F$_3$CBr$\cdots$OCH$_2$, PheBr$\cdots$SHCH$_3$, and F$_3$CBr$\cdots$C$_6$H$_6$, respectively), unlike in the case of AXZP basis sets, which have even slightly more basis functions (by about 4 -- 7\%) than aug-cc-pVXZ basis sets. Concerning the ways of SREs inclusion, the use of pseudopotential appears to be as good choice as the use of DKH Hamiltonian, as their mutual difference in interaction energies ranges between 0.03 -- 0.06 kcal/mol.

Detailed analysis of the $\Delta$Es for iodine-containing species (Table S3) is rather restricted, due to the aforementioned limited availability of systematically optimized basis set series for iodine. In this case, the most accurate $\Delta$Es with SREs inclusion using DKH Hamiltonian are obtained in ANO-RCC basis sets, which are notably less saturated compared to aug-cc-pVXZ basis sets with pseudopotentials. The differences between these two sets of results are, however, quite small ranging from zero for F$_3$CI$\cdots$OCH$_2$ to 0.14 kcal/mol for F$_3$CI$\cdots$C$_6$H$_6$. This comparison is even more problematic due to the fact, that ANO-RCC results are obtained using full CCSD(T) calculations, while the results in aug-cc-pVXZ-PP series are obtained using focal-point scheme potentially introducing additional uncertainty. Nevertheless, the deviation from the most accurate nonrelativistic results (obtained using AXZP basis sets) can still be identified, particularly in the case of PheI$\cdots$SHCH$_3$, where it reaches the maximum of 0.45 kcal/mol (about 16\%, decreases from 0.73 at double-$\zeta$ to 0.51 kcal/mol at quadruple-$\zeta$ level).

\subsection{Bond Critical Points}

\begin{figure}
    \centering
    \includegraphics{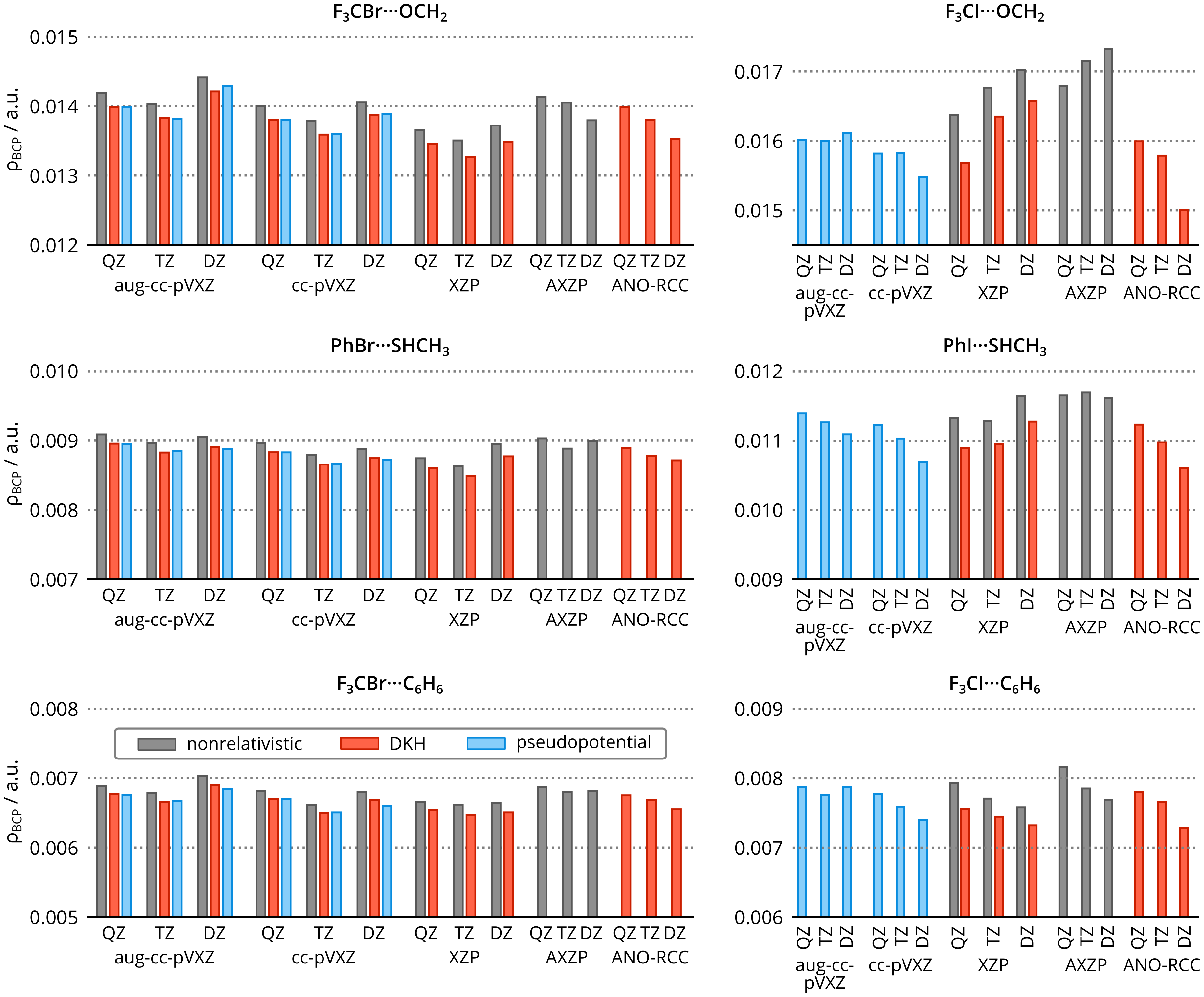}
    \caption{Electron densities $\rho_{BPC}$ at the bond critical points with nonrelativistic (gray), DKH-Hamiltonian optimized (red) and pseudopotential (blue) basis sets using frozen-core approximation. Note the common Y-axis extent for all of the complexes ($3\cdot10^{-3}~\mathrm{a.u.}$).}
    \label{fig:bcps}
\end{figure}

The values of electron density at BPCs are shown in \Cref{fig:bcps}. They span a range between 0.006 and 0.017 a.u. $\rho_\mathrm{BCP}$ is slightly sensitive to the basis set quality and size as well as to the treatment of SREs. The largest variations are observed for F$_3$CI$\cdots$OCH$_2$ complex. In all cases, where it was possible to compare, the relativistic $\rho_\mathrm{BPC}$ is smaller than the nonrelativistic. The difference is about 1\% for complexes containing bromine, and about 4\% for complexes containing iodine.

\begin{figure}
    \centering
    \includegraphics{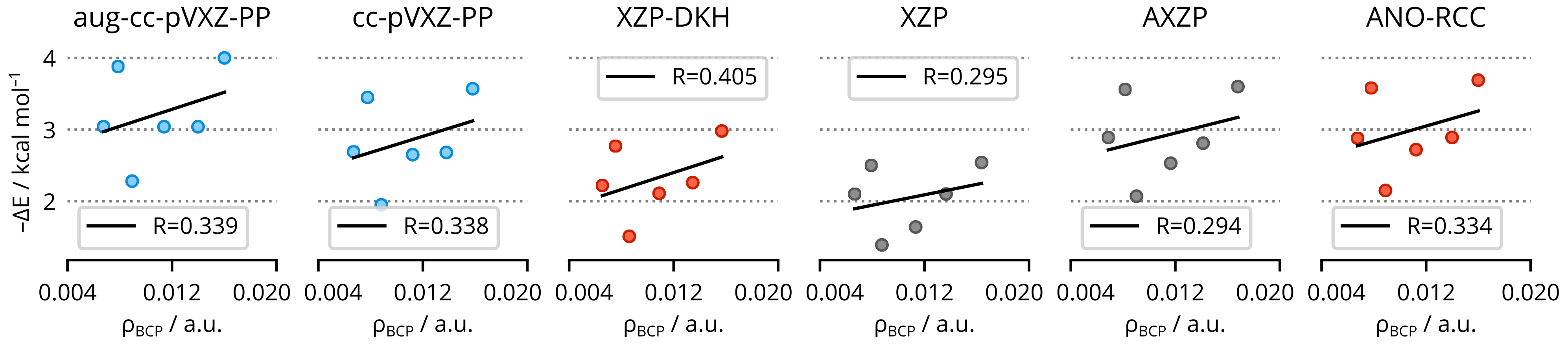}
    \caption{A correlation between the electron density at the bond critical points ($\rho_\mathrm{BCP}$ and the interaction energy ($\Delta E$). The scatter points are color coded according to the SRE treatment, \eg nonrelativistic in gray, relativistic pseudopotential in blue, and relativistic DKH in red.} 
    \label{fig:correlation}
\end{figure}

Previously, it was shown that $\rho_\mathrm{BCP}$ is a measure of interaction strength of noncovalent complexes \cite{Grabowski04, Bankiewicz12, Franconetti19}, so we analyzed the 432 pairs of $\rho_\mathrm{BCP}$ and $\Delta E$ collected here. \Cref{fig:correlation} shows the regression plots for the quadruple-$\zeta$ basis sets common for brominated and iodinated complexes. Bearing in mind the small number of complexes, we observe only a poor correlation. Here the complexes are more diverse and span a smaller range of interaction energies than in previous studies, where systematic sets of complexes shown higher correlations \cite{Amezaga10, Bauza11}. Overall, a wider set to complexes would be needed to draw any conclusions about the role of SREs on the correlation between $\rho_\mathrm{BCP}$ and $\Delta E$.

\section{Conclusions}

In this work, we provide a detailed analysis of the impact of scalar relativistic effects on properties of \shole~and of X-bonds. Using  CCSD(T)/CSB -- the golden standard of noncovalent calculations -- we eliminate inaccuracies related to the electron correlation. By carrying out calculations in a series of systematically optimized basis sets we detach the effect of basis set incompleteness from SREs, which turned out to be a challenging task, in particular for iodine-containing species. 

In general, the SREs are small for all investigated quantities of halogenated monomers and X-bonded complexes. The magnitude of SREs is intuitively larger for iodine than for bromine. For brominated complexes, the SREs are mostly negligible ($<$5\%). SRE treatment is essential for iodinated molecules/complexes, where the error brought by the neglect of SREs may reach several tens of percent.

We found that the magnitude of basis set effects on interaction energies and \shole ~properties is on par to or even larger than SREs. The proper choice of basis set is thus as critical as the ways of SRE treatment. Smaller double$-\zeta$ basis sets sometimes performed better than triple-$\zeta$ basis sets perhaps due to error compensation, which increases the motivation to use extrapolation to CBS.

Calculations with relativistic pseudopotentials delivered results similar to those from all-electron calculations using relativistic Hamiltonian.

\section*{Funding Information}

This work was supported by the Slovak Research and Development Agency, contract No. APVV-15-0105, and Slovak Grant Agency VEGA under the contract no. 1/0777/19.

\section*{Research Resources}

Part of the calculations was performed in the Computing Center of the Slovak Academy of Sciences using the supercomputing infrastructure acquired in project ITMS 26230120002 and 26210120002 (Slovak infrastructure for high-performance computing) supported by the Research \& Development Operational Programme funded by the ERDF.

\section*{Author Contributions}

MP designed the research. DS and MP performed quantum chemical calculations. MHK evaluated monomer properties. All of the authors evaluated properties of noncovalent complexes and drafted the manuscript. MHK and MP finalized the manuscript.

\section*{Supplementary Information}

The number of eliminated basis set functions due to linear dependence, and the interaction energies of all complexes at all levels of theory are provided in Tables S1, S2 and S3.

\clearpage
\newpage

\providecommand{\latin}[1]{#1}
\makeatletter
\providecommand{\doi}
  {\begingroup\let\do\@makeother\dospecials
  \catcode`\{=1 \catcode`\}=2 \doi@aux}
\providecommand{\doi@aux}[1]{\endgroup\texttt{#1}}
\makeatother
\providecommand*\mcitethebibliography{\thebibliography}
\csname @ifundefined\endcsname{endmcitethebibliography}
  {\let\endmcitethebibliography\endthebibliography}{}

\newpage
\clearpage

\renewcommand\thefigure{S\arabic{figure}}
\renewcommand\thetable{S\arabic{table}}
\renewcommand\theequation{S\arabic{equation}}
\setcounter{figure}{0}
\setcounter{table}{0}
\setcounter{equation}{0}

\section*{Supplementary Information}

From the main text of the article we repeat that the energies of complexes and their interacting subsystems were calculated using the CCSD(T) method, wherever it was computationally feasible. The calculations in AQZP and aug-cc-pVQZ basis set of all complexes except for the smallest one -- F$_3$CBr$\cdots$OCH$_2$, could only be carried out at the MP2, and the well-established focal-point scheme was applied to obtain an estimate of the $E_Q$(CCSD(T)) energies.
\begin{equation}
\label{eq:focalsi}
    E_X(\mathrm{CCSD(T)}) =  E_X(\mathrm{MP2}) + \Delta E_{X-1}(\mathrm{CCSD(T)})
\end{equation}
where $\Delta E_{X-1}$(CCSD(T)) is the difference of CCSD(T) and MP2 energies in basis set with lower cardinality ``X -- 1''.

\begin{table}
\caption{Number of eliminated basis functions due to linear dependence / total number of AO basis functions for complexes containing iodine.}
\label{tabSI:thresholds}
\begin{center}
\begin{tabular}{lr|ccc}
\hline \hline
Complex                      & Basis set      & X=D & X=T & X=Q \\ \cr

 F$_3$CI$\cdots$OCH$_2$
                             & AXZP           &  &  &  \\
                             & ANO-RCC-X      &  &  &  \\
                             & cc-pVXZ-PP     &  &  &  \\
                             & aug-cc-pVXZ-PP &  &  &  \\ \cr
               
 PheI$\cdots$SHCH$_3$
                             & AXZP           & 2/346 & 10/687 & 203$^{(1)}$/1216 \\
                             & ANO-RCC-X      &   -   &    -   & 2/742 \\
                             & cc-pVXZ-PP     &   -   &    -   & 1/781 \\
                             & aug-cc-pVXZ-PP & 3/301 & 12/634 & 24/1150 \\ \cr
F$_3$CI$\cdots$C$_6$H$_6$ 
                             & AXZP           & 2/358 & 11/703 & 33$^{(2)}$/1231 \\
                             & ANO-RCC-X      &    -  &     -  &  3/740  \\
                             & cc-pVXZ-PP     &    -  &     -  &  1/797  \\
                             & aug-cc-pVXZ-PP & 3/316 & 11/653 & 23/1168 \\ \cr

\hline\hline
\end{tabular}
\begin{flushleft}
linear dependence threshold [a.u.]: $^{(1)}$ 7,5.10$^{-3}$; $^{(2)}$ 3,0.10$^{-5}$
\end{flushleft}
\end{center}
\end{table}

\begin{table}[tb!]
\caption{CCSD(T) interaction energies [kcal/mol] of complexes containing bromine. Interaction energies in parentheses are obtained using focal-point approximation according to Eq.~\ref{eq:focalsi}.}
\label{tabSI:eint_br}
\begin{center}
\begin{tabular}{lrc|llll}
\hline \hline
Complex             & Basis set & CBS & X=D & X=T & X=Q \\ \cr

F$_3$CBr$\cdots$OCH$_2$
                        & XZP             & --2.65 (--2.78) & --0.63 & --1.35 & --2.10 \\
                        & cc-pVXZ         & --2.92 (--2.79) & --1.14 & --1.92 & --2.50 \\ 
                        & AXZP            & --3.04 (--2.98) & --2.21 & --2.57 & --2.81 \\ 
                        & aug-cc-pVXZ     & --3.00 (--2.99) & --2.37 & --2.69 & --2.87 \\
                        & XZP-DKH         & --2.79 (--2.92) & --0.77 & --1.52 & --2.26 \\ 
                        & cc-pVXZ-DKH     & --3.06 (--2.93) & --1.30 & --2.07 & --2.64 \\
                        & aug-cc-pVXZ-DKH & --3.13 (--3.13) & --2.53 & --2.83 & --3.01 \\
                        & ANO-RCC-X       & --3.14 (--3.15) & --1.80 & --2.54 & --2.89 \\
                        & cc-pVXZ-PP      & --3.10 (--2.97) & --1.36 & --2.10 & --2.68 \\
                        & aug-cc-pVXZ-PP  & --3.17 (--3.17) & --2.54 & --2.86 & --3.04 \\ \cr
               
PheBr$\cdots$SHCH$_3$
                        & XZP             & --1.94 & --0.05 & --0.63 & --1.39 \\
                        & cc-pVXZ         & --2.24 & --0.46 & --1.22 & --1.81 \\
                        & AXZP            & --2.22 & --1.38 & --1.87$^{(a)}$ & --2.07$^{(a)}$ \\
                        & XZP-DKH         & --2.06 & --0.02 & --0.77 & --1.51 \\ 
                        & aug-cc-pVXZ     & --2.25 & --1.59 & --1.98$^{(a)}$ & --2.13$^{(a)}$ \\
                        & cc-pVXZ-DKH     & --2.35 & --0.60 & --1.34 & --1.92 \\
                        & aug-cc-pVXZ-DKH & --2.36 & --1.72 & --2.09$^{(a)}$ & --2.25$^{(a)}$ \\
                        & ANO-RCC-X       & --2.41 & --0.72 & --1.79 & --2.15 \\
                        & cc-pVXZ-PP      & --2.38 & --0.62 & --1.37 & --1.95 \\
                        & aug-cc-pVXZ-PP  & --2.39 & --1.73 & --2.13$^{(a)}$ & --2.28$^{(a)}$ \\ \cr
               
F$_3$CBr$\cdots$C$_6$H$_6$
                         & XZP             & --2.84 & --0.01 & --1.10 & --2.10 \\
                         & cc-pVXZ         & --3.03 & --0.83 & --1.96 & --2.58 \\ 
                         & AXZP            & --2.99 & --2.13 & --2.76$^{(a)}$ & --2.89$^{(a)}$ \\
                         & aug-cc-pVXZ     & --3.02 & --2.34 & --2.80$^{(a)}$ & --2.93$^{(a)}$ \\
                         & XZP-DKH         & --2.96 & --0.08 & --1.20 & --2.22 \\ 
                         & cc-pVXZ-DKH     & --3.11 & --0.94 & --2.06 & --2.67 \\
                         & aug-cc-pVXZ-DKH & --3.19 & --2.43 & --2.89$^{(a)}$ & --3.06$^{(a)}$ \\
                         & ANO-RCC-X       & --3.13 & --1.59 & --2.55 & --2.88 \\
                         & cc-pVXZ-PP      & --3.13 & --0.94 & --2.09 & --2.69 \\
                         & aug-cc-pVXZ-PP  & --3.13 & --2.44 & --2.91$^{(a)}$ & --3.04$^{(a)}$ \\ \cr

\hline\hline
\end{tabular}
\begin{flushleft}
$^{(a)}$ Energy calculated according to Eq.~\ref{eq:focalsi}.
\end{flushleft}
\end{center}
\end{table}

\clearpage

\begin{table}[tb!]
\caption{CCSD(T) interaction energies [kcal/mol] of complexes containing iodine.}
\label{tabSI:eint_i}
\begin{center}
\begin{tabular}{lrc|llll}
\hline \hline
Complex                      & Basis set      & CBS   & X=D & X=T & X=Q \\ \cr

 F$_3$CI$\cdots$OCH$_2$
                             & XZP            & --2.86 & --0.98 & --2.08 & --2.54 \\
                             & AXZP           & --4.09 & --2.41 & --2.93 & --3.60 \\
                             & ANO-RCC-X      & --4.19 & --2.20 & --3.00 & --3.69 \\
                             & XZP-DKH        & --3.42 & --1.34 & --2.38 & --2.98 \\
                             & cc-pVXZ-PP     & --4.09 & --1.97 & --2.85 & --3.57 \\
                             & aug-cc-pVXZ-PP & --4.19 & --3.33 & --3.75 & --4.00 \\ \cr
               
 PheI$\cdots$SHCH$_3$
                             & XZP            & --1.99 & --0.89 & --1.16 & --1.64 \\
                             & AXZP           & --2.75 & --1.54 & --2.22$^{(a)}$ & --2.53$^{(a)}$ \\
                             & ANO-RCC-X      & --3.27 & --0.66 & --1.97 & --2.72 \\
                             & XZP-DKH        & --2.58 & --0.41 & --1.48 & --2.11 \\
                             & cc-pVXZ-PP     & --3.19 & --0.97 & --1.91 & --2.65 \\
                             & aug-cc-pVXZ-PP & --3.20 & --2.27 & --2.82$^{(a)}$ & --3.04$^{(a)}$ \\ \cr
               
 F$_3$CI$\cdots$C$_6$H$_6$ 
                             & XZP            & --2.85 & --0.24 & --2.02 & --2.50 \\
                             & AXZP           & --3.65 & --2.54 & --3.44$^{(a)}$ & --3.56$^{(a)}$ \\
                             & ANO-RCC-X      & --4.14 & --1.58 & --2.82 & --3.58 \\
                             & XZP-DKH        & --3.15 & --0.55 & --2.25 & --2.77 \\
                             & cc-pVXZ-PP     & --3.98 & --1.31 & --2.72 & --3.45 \\
                             & aug-cc-pVXZ-PP & --4.00 & --3.02 & --3.71$^{(a)}$ & --3.88$^{(a)}$ \\ \cr

\hline\hline
\end{tabular}
\begin{flushleft}
$^{(a)}$ Energy calculated according to Eq.~\ref{eq:focalsi}
\end{flushleft}
\end{center}
\end{table}

\clearpage

\end{document}